\documentclass[runningheads,a4paper]{llncs}

\usepackage{amssymb}
\usepackage{graphicx}

\usepackage{url}
\urldef{\mailsa}\path|{alfred.hofmann, ursula.barth, ingrid.haas, frank.holzwarth,|
\urldef{\mailsb}\path|anna.kramer, leonie.kunz, christine.reiss, nicole.sator,|
\urldef{\mailsc}\path|erika.siebert-cole, peter.strasser, lncs}@springer.com|

\begin{document}

\mainmatter  

\title{Task Allocation in Robotic Swarms: Explicit Communication Based Approaches
}

\titlerunning{Task Allocation in Robotic Swarms}

\author{Aryo Jamshidpey\inst{1} \and Mohsen Afsharchi\inst{2}} 
\institute{Institute for Advanced Studies in Basic Sciences\\
\email{aryo.jamshidpey@iasbs.ac.ir}
\and Department of Computer Science, University of Zanjan\\
\email{afsharchim@znu.ac.ir}}

\toctitle{Lecture Notes in Computer Science}
\tocauthor{Authors' Instructions}
\maketitle

\begin{abstract}
In this paper we study multi robot cooperative task allocation issue in a situation where a swarm of robots is deployed in a confined unknown environment where the number of colored spots which represent tasks and the ratios of them are unknown. The robots should cover this spots as far as possible to do cleaning and sampling actions desirably. It means that they should discover the spots cooperatively and spread proportional to the spots area and avoid from remaining idle. We proposed 4 self-organized distributed methods which are called hybrid methods for coping with this scenario. In two different experiments the performance of the methods is analyzed. We compared them with each other and investigated their scalability and robustness in term of single point of failure.
\end{abstract}
\section{Introduction}

Swarm robotics is a branch of robotic science which is inspired by social insects and other nature colonies that show complex behavior although they have simple members. Defining the roles of robots in society is one of the most important problems for simulating such behaviors. The action of assigning tasks to agents for performing is called task allocation. Tasks are dependent to global mission. If robots perform their tasks effectively this mission is performed effectively too. From a control architectural perspective Burger \cite{3} distinguishes between Heteronomous, Autonomous and Hybrid methods in task allocation. In this paper, we introduce a practical scenario for the issue of task allocation in swarm robotics and 4 hybrid methods for solving it in unknown environments which, the number, locations and ratios of tasks are unknown to robots.

Market-based mechanism is one of the main approaches that tackle the task allocation problem. TraderBots is one of the works in this subject that is presented by Dias \cite{6}. In TraderBots the robots bid on tasks on the basis of cost and it is given to a bidder with lowest cost. A comprehensive study of market-based multi-robot coordination can be found in \cite{7}. Most solutions in self-organized task allocation is threshold-based that are inspired by models initially proposed to describe the behavior of insect societies \cite{1}. In this case we can mention Krieger and Billeter work \cite{9} which benefits from a simple threshold-based model for task allocation in a foraging scenario.Labella et al. \cite{10} and Lui et al. \cite{11,12} proposed two probabilistic task allocation approaches which use adaptive thresholds.Brutschy et al. in\cite{2} presented a task allocation strategy in which robots specialize to perform tasks in the environment in a self-organized. Jones and Mataric \cite{8} introduced an adaptive distributed autonomous task allocation method for identical robots. In this work a task allocation is called desirable in which the ratio of robots that doing the same kind of task equal with the ratio of that task in the environment. Dahl et al. \cite{4} proposed a method that controls the group dynamics of task allocation. In this work each robot can choose between two separated foraging cycles. Nouyan et al \cite{15} presented autonomous rule-based task allocation method in which robots attempting to identify and transfer a food to the nest by using a set of rules and forming a chain in a self-organized way. Dasgupta \cite{5} presented a communication-based method for task allocation. Each task needs multiple robots to be done. Robots can only partially complete tasks and one after the other contribute to progressing them. Our proposed practical scenario like Dasgupta's one is about unknown environments. The discussion part is dedicated for comparing these scenarios and their solutions and also stating their differences.
\section{Problem Definition}
This scenario involves a colony of identical robots with limited energy levels that are rechargeable and an environment full of obstacles and colored spots which represent types of tasks.The individuals are unaware of the size of the population and the distribution of other robots.At any moment, each robot is able to do only one of the “forage for green” or “forage for black” subtasks.Depending on the area of the spots, the number of robots to do cleaning and sampling actions in them varies. Obviously it is not necessary to fill the whole capacity of spots with robots. Even a robot is able to do cleaning and sampling actions on its own but it takes much time and less desirable.In abstract the scenario can be defined in terms of finding more colored spots with minimal energy waste. One of the causes of this waste is unnecessary robot turns in the environment.Also there are some other reasons for energy waste such as maximum spreading of robots in different spots proportional to their area, avoiding robots from collision with obstacles and finally preventing active robots from remaining idle.
\section{Methods}
In all proposed methods the robots are initially in random places. In each method the energy level of each robot is divided into three parts. First part is for foraging. Second part is to do cleaning and sampling actions and the third part is dedicated for returning to the charging station, either nest or any other specified location. All methods are organized based on transferring messages and all messages have the same structure. Each message may have several rows that each of them presents a distinct spot’s information. Information of each spot consists of 10 features which are as follows: A row is occupied or not, that is shown by 1 or 0 similarly. X, Y and Z coordinate of the current position of the robot in the desired environment, which is calculated via GPS. The total capacity of the existing spots (Number of robots required to do cleaning and sampling actions in spots, which vary depending on the area of the discovered spot). The current number of robots deployed in the desired spot. The number of robots needed to complete the spot capacity. The number of hops (In order to estimate the number of robots that have been informed of the discovered spot and to identify the spots which expose to saturation and starvation). Time elapsed from the beginning of the running of the code and finally color of the discovered spot which are conventionally denoted by numbers (for example in our simple scenario black and green are denoted by 0 and 1 respectively).

Message updating procedure is the same in all of proposed methods. Every robot has a message that contains information about the spots, called private message. This message is empty at first and then updated via occurrence of two different events. First when a robot finds a non-observed spot for the first time, and second when it receives a message from one of its neighbors:

1.when a robot explores a non-seen spot for the first time, it starts to calculate the center of spot and also its area, computes its capacity based on the area and calculate other features of message based on a pre-defined rule. It should be considered that if a robot entered a spot under the guidance of another robot, it would subtract the number of required robot field by one, add one unit to the number of existing robots field and then set the current time as the row time field.

2. In this situation two different tasks may exist for the receiver to do, one is to add non existing rows to its message. The order of the updated message is the same as sender's one and in every new row, hop count will be increased by one. The other task is to replace existing rows according to the time field. If the row associated with a specific spot that exists in both private and received messages has greater time field in the received message, then it will be replaced in the private message.
\subsection{Static communication based method}
This method is a hybrid method combining the autonomous and distributed methods \cite{3}. At the beginning, one of the two subtasks of exploring for green or black spot is assigned to each robot which is called worker statically (With a probability of 0.5). Static means that the assigned task will not change unless the robot runs out of its searching energy.Each robot moves in the environment by random walk and prevents collision with other robots, obstacles and walls by the help of its distance computing sensors. Each robot updates its private message periodically and floods it to every other robot in the coverage area of its radio frequency transmitter. By this process, messages will propagate among the robots (either deployed or searcher robots). If a robot could not find a match (a spot with the same color of its associated subtask color) before running out its searching energy, it would go to one of the spots in its private message. By having this policy, our goals for preventing idle robots existence, maximum spot coverage and also preventing spot starvation will be guaranteed in a desirable level.

 The use of desirable word is due to the energy limitation and also the heuristic sense of the scenario. Therefore we proposed a method for trap prevention by partitioning robot energy and then searching the environment for the best spot while some amount of energy exists and also the robot reaches the best possible spot (the first seen spot by a robot is not necessarily the best spot for it, so it will search for a suitable spot until it runs out of searching energy). At the end the robot reaches the decision step that is as follow:
\subsubsection{Appropriate spot selection decision making.}
One of the advantages of the proposed method is the way it decides to walk to the most appropriate spot. As the searching energy of one robot reaches its threshold, the robot starts the decision making step. At this time the robot should sort its private message based on three fields; color, hop count and its current distance from the spots. At first this sort is done based on color field, so that spots with the same color as the robot current state will be placed at the top. Now the sorted list has two parts, one for the same colors as the subtask color and another for opposite colors to it. Then these lists are sorted based on the hop field. Each sorted list then will be categorized according to sets of 5 hops (number of hops per category should be set based on the spots area and count in the environment). The reason for using this parameter is to prevent spot starvation and saturation. When a robot finds out that the hop count of a spot is large, it estimates that at most one less than the hop count robots are aware and can reach the spot before it. Hence the robot considers the spot may get saturated. Robots prefer to go to spots with low hop number to prevent starvation. In the third step the sorted list based on color and hop count will be further sorted based on Euclidean distance. So in each hop category, spots are sorted based on their distance from the robot.

If a robot in its decision step observes a spot with the same color of its assigned color while transferring to its appropriate spot, it will explore the spots features. If the spot is suitable for the robot and if saturation will not occur, current spot will be accepted by the robot and necessary updates will be done. It is possible that before a robot reaches to its highest priority spot, the capacity of that becomes full. In this case the robot will be aware of this occurrence by the messages received from other deployed robots in that spot. It will leave the spot and will begin to make another decision from its private list. We can consider an integer threshold in such cases. Robots will try to make decision until the threshold. If a robot cannot find a suitable spot in its decision step before the threshold, it will return to home or the nearest charge station to be recharged and then starts to search again. The state diagram of the robot’s controller while using this method is shown in figure 1.
\subsection{Dynamic communication based method}
This method (figure 1) is similar to the first method in a way that it supports dynamic task allocation in such a way that the robot will change its target color probabilistically after a time step. For example consider that a robot has 4 spots in its private list after 100 iterations containing 1 black and 3 green. Then the robot will set its target color to black by probability ¼ and also will set it to green by probability ¾. This is why we expect this method to be better than the previous one, because in this method the searching time for each robot to find the appropriate spot may be decreased. if the robots are not distributed uniformly in the environment at the beginning, it is possible that some spots are not discovered after a small time steps (100 time steps). This is due to their large distances from the initial places of the robots and also attitude changing mechanism of the robots.
\begin{figure}[top]
\centering
\includegraphics[height=5cm]{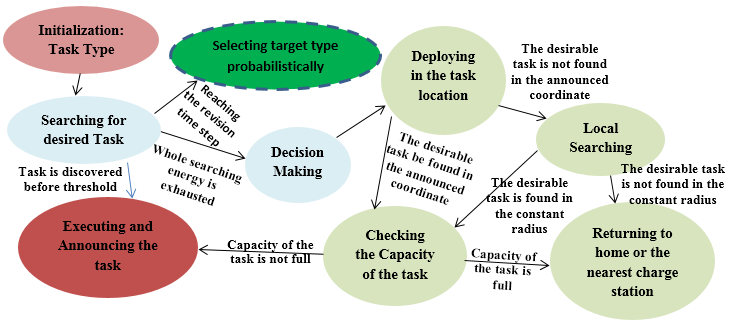}
\caption{Method1 and Method 2 (by considering the dark green state which is shown by dashed border) state diagram.}
\label{fig:example}
\end{figure}
\subsection{Decentralized Chapar Method}
In this method we have some radio turrets called Chapar stations which are used for radio communications and also we assume two groups of robots; workers (as defined in the previously explained methods) and Chapars that transfer messages between workers and Chapar stations. "Chapar Khaneh" or "Chapar-Khaneh" is a term in Persian, meaning the "house of courier" as "Chapar" means "courier", referring to the postal service used during the Achaemenid era \cite{16}. Here we used Chapar station instead of Chapar-khaneh.

High speed robots with simple structure are considered as Chapars. Chapars flood their updated private message that is modified by some workers to the area of their radio frequency coverage. Once a Chapar realizes a new row in its private message (showing that a new spot has been discovered) it  quickly goes to the nearest Chapar station and sends its private message to it and also updates the message based on the content of the messages sent by the Chapar stations. In addition each Chapar goes to the nearest Chapar station periodically (e.g. each 100 time steps) to update its private message and access to the most recently information about the status of the spots.

Chapar stations are also in the coverage of each other and so they replicate messages to keep the whole system up to date. In the simplest variation of this model, there could be only one Chapar station whit limited coverage area. To achieve high performance with the lowest cost, this method needs to coordinate Chapar stations in a manner that their communication radiuses cover each other sequentially. Although this method has a higher message transfer speed with respect to the first and second methods, but it needs radio frequency transmitters with higher coverage area and also in large environments it needs more Chapar stations. Nevertheless, this method can perform fine even if Chapar stations are not in the coverage area of each other. In this situation the messages may not be similar in all Chapar stations but we can assume the environment as some sub-environments each with a single Chapar station that can transform messages with high speed. In a timely manner the information of one area may be transferred to another area by Chapars. The state diagram of the Chapar’s controller while using this method is shown in figure 2.
\subsection{Centralized Chapar Method}
\begin{figure}[bottom]
\centering
\includegraphics[scale=0.50]{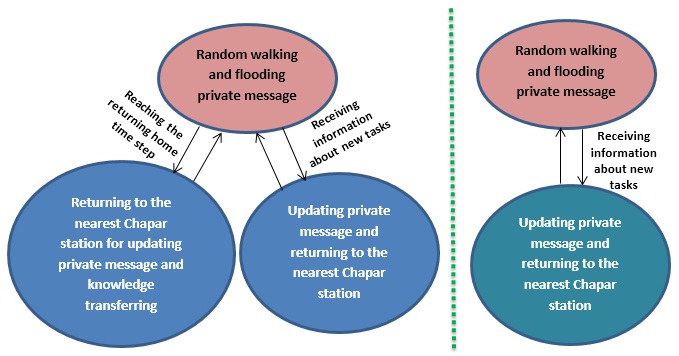}
\caption{Left: Method 4 State Diagram, Right: Method 3 State dDiagram.}
\label{fig:example}
\end{figure}
This method (figure 2) is similar to the decentralized one in which there is only one Chapar station that covers the entire environment. In this, method Chapars have a single task to transfer messages from robots to the Chapar station and they do it once they realize a new row in their private messages. When the Chapar station receives a message that is not similar to its private message, it updates the private message and sends it to all robots.
To prevent the Chapar station from single point of failure, in addition to message updating by Chapars, they should transfer message in the environment quickly. In spite of the centralized method, each Chapars only goes to the Chapar station when it encounters a new spot in its private message.
It should be reminded that the robots used in the third and fourth methods, have the same behavior as the robots that were used for one of the first or second method (optionally) and as explained before, their behavior are the same as one of them.
\section{Simulations}
We have used e-puck robots in simulations which are extended by color detection Infrared sensors, positioning module, compass and limited-range radio waves emitter and receiver. Since the purpose of this article is to involve a wide range of robots and the use of simpler hardware, we considered them without any camera. In this way the methods can show their power in using of blind robots. All experiments have been implemented in a 3m x 3m square environment enclosed with walls by using of Webots as robotic simulation software. In two general experiments, performance of the four proposed methods is evaluated individually before and after energy consumption and in both of them 10 robots which are called workers with IDs from 1 to 10 are used. Initially foraging mode of the robots with IDs from 1 to 5 is adjusted to green and the robots with IDs from 6 to 10 is adjusted to black. 

Further more, the speed of each method is determined by its foraging(searching) energy threshold, hence setting the threshold is very important. It should not be so low that the robot does not have enough time for foraging and also it should not be so high that robots use their whole energy without any success or much of it which in this case its performance will be reduced when it deploys in a spot. Some factors such as area of environment, ratios of tasks, density of tasks and the number of robots can help us to determine an appropriate value for this parameter. In all of our experiments, the energy threshold level is considered as 250 iterates of each robot’s control code.

For simplicity, the details of cleaning and sampling operations are ignored, the colored spots are considered as 30cm x 30cm squares and finally the length of each robot’s communication radius is considered larger than the diameter of each colored square. In both experiments, for each 300 square centimeters of each spot one worker is sufficient for covering it desirably, which means that all spots are covered desirably with 3 workers. Indeed as mentioned previously, with less number of workers spot covering is also possible, but with lower speed.

\subsection{First experiment: performance evaluation of proposed methods before the threshold energy}
Since the third and fourth methods use the worker robot which its controller is that of the first or second method, it is not necessary to compare the performance of all methods before foraging energy consumption (before the threshold).So we have compared only the performance of the first and second methods before the threshold. For this purpose four environments covered by green and black spots and also containing obstacles are considered. The first environment has 3 green spots and 3 black ones, the second has 4 green spots and 2 black ones, the third includes 5 green spots and 1 black ones and finally the fourth contains only 6 green spots.
In each of the four areas, both the first and second methods are tested 10 times separately. The average number of successful robots before the energy threshold for the fist and second methods is shown in figure 3 (right).

It can be seen that, except for the first environment in which the average number of successful robots before the threshold are equal for both methods, in other environments, the average number of successful robots in the second method is higher than the first one. This disparity grows by moving from the first environment to the fourth and its reason is the changing attitude mechanism which is used by second method's robots during their foraging operation. As a result this leads to increasing in the number of robots which their attitude changes when the number of green spots rise and the number of black ones falls respectively. Subsequently this process will result in forming approximate stability in the number of successful robots before the threshold. As it is shown in figure 3 (right), we can conclude that in unknown environments where the number of colored spots (black and green in our example) and the ratios of them are unknown, the second method is more successful than the first one in terms of the robots' attempts in finding spots by themselves before energy threshold.

In the next step the average number of green spots that have been found by workers(by both the first and the second methods) before the threshold is shown in figure 3 (left). In this figure the obtained curve from the second method is steeper than the first method’s one. Its primary reason is increasing the number of green spots and decreasing the black ones which results in increasing the probability of finding new green spots in both methods. But the reason that the second method’s curve is more steeper, is increasing of the number of workers which search for green spots during the search time. As mentioned before this is because of changing the robot’s attitudes during foraging operations.
\begin{figure}[top]
\centering
\includegraphics[height=4.5cm, width=14.3cm]{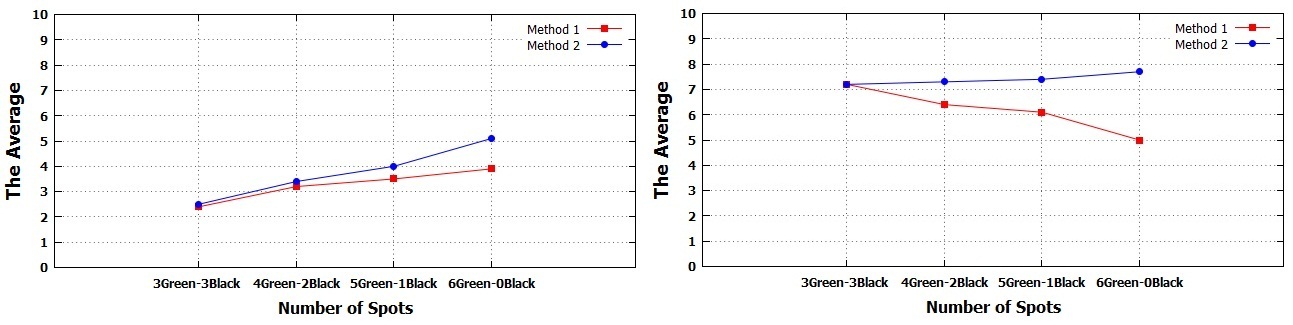}
\caption{Right: Average number of successful robots before the threshold. Left:Average number of discovered spots.}
\label{fig:example}
\end{figure}
Figure 3 (left) shows that in unknown environments where the number of colored spots (black and green in our example) and their ratios are unknown, the second method is more successful than the first one in spot finding before the energy threshold.

Figure 4 (right) shows the average number of robots deployed in the green spots before the end of the foraging energy in both the first and second methods. It can be observed that the steepness of the second method’s curve is ascending linear but the steepness of the first method’s curve is sub linear. In the first method the maximum number of robots in green spots is 5 because only 5 robots have green initial foraging modes and there is no any changing attitude mechanism before threshold. But in the second method there is changing attitude mechanism and so the maximum number of robots in green spots might be 10 and this means the whole colony. We can conclude that in unknown environments where the number of colored spots (black and green in our example) and the ratios of them are unknown, adaptability has a significant positive impact on the performance of robots. As it can be observed, the second method is more successful than the first one in the average number of robots which have deployed in the green spots before the threshold. To sum up, from the above results in unknown environments with the features which are mentioned above, the second method is more efficient than the first one before the energy threshold.
\subsection{Second experiment: performance evaluation of proposed methods after the threshold energy}
We use an environment consisting of 3 green spots and 3 black ones for evaluating the performance of the methods after foraging energy consumption. In this area, all proposed methods are separately tested 10 times with random initial distribution of robots. In the third and fourth methods, in addition to 10 workers, another 3 Chapar robots that are faster than the workers are used too. It should be mentioned about Chapar station that the third method is equipped with 3 Chapar stations which their communication radius cover each other sequentially and the fourth method is equipped with one of them, which is omniscient. Figure 4 (left) shows the results of the second experiment in abstract. The absorption percentage is the percentage of successfulness of finding spots by robots after the threshold by applying the decision making mechanism.

As expected, the fourth method has the highest absorption percentage which means 100\%. This is due to the use of global message transferring system. The third method is in second place with 87.09\% and after it the first and the second methods with approximate absorption 76\% are both in third place. It should be reminded that both of the first and second methods have the same communication mechanism. Higher speed in communicating will lead to making a better task selection based on their colors, distances and number of hops by unsuccessful workers. It results in a better robot distribution that reduces saturation and starvation as far as possible. It also causes more unsuccessful robots attract to the spots.

\section{Discussion}
Most studies in swarm robotic define their own test scenario, which is then used to build a concrete swarm robotic system capable of solving the problem. This leads to a huge amount of differently designed global missions and as a result to many different solutions which are hard to compare\cite{3}. Thus in most of the proposed methods in this area, researchers have only introduced their own methods and refrained from comparing with other methods. Our scenario also possesses different features, goals and finally distinct global foraging mission compared to previous scenarios in task allocation field. Accordingly applying current approaches in our scenario and consequently comparing them with each other is so difficult and in most of the times is impossible, except for changing the scenario which in turn leads to changing both the problem and the solutions. As pointed out before, Dasgupta`s method is practical for unknown environments with this difference that its global mission is different from what have been proposed here. However our proposed methods have distinct advantageous over Dasgupta's. 

The proposed methods have the least waste time for robots as apposed to that of Dasgupta in which, robots might be idle in environment for a long time. These methods have been designed for the energy constrained scenario compared to Dasgupta's. Consequently the priority of discovering new spots is higher than accomplishing a task and then continuing unlimited foraging. Further more in contrast with Dasgupta's method(which robots may be in idle mode for a long time) our methods are close to optimal in the sense of idle mode.

On the other hand, our approaches are very practical for scenarios which detecting a task should be done in a limited time. However in Dasgupta's approach, since after detecting a task other neighbor robots quit foraging and wait for accomplishing the task sequentially, the probability of identifying other tasks in a limited time will decrease. Moreover the proposed methods in this paper are practical for blind robots while Dasgupta's method is practical for robots with cameras which ignoring use of them obviously reduces cost in swarm robotics. Also there is no more need to apply sophisticated techniques of machine vision in order to determine starvation (which may have some errors).
\subsection{Scalability and Robustness in term of single point of failure}
In this section, we will analyze the scalability and robustness of the proposed methods with respect to their efficiency. If by adding or removing robots the performance of a method drops off, we will call it unscalable. Accordingly, both the first and second methods are scalable because both are distributed, autonomous and based on the local communication. So Adding or removing robots will maintain their effectiveness. The decentralized Chapar method also has scalability property. This is due to using of the limited range Chapar station, the worker robots and the Chapar robots that are distributed and behave locally. But the centralized Chapar method, due to using of omniscient Chapar station with respect to its expected efficiency does not have scalability property.
\begin{figure}[top]
\centering
\includegraphics[height=5cm, width=14cm]{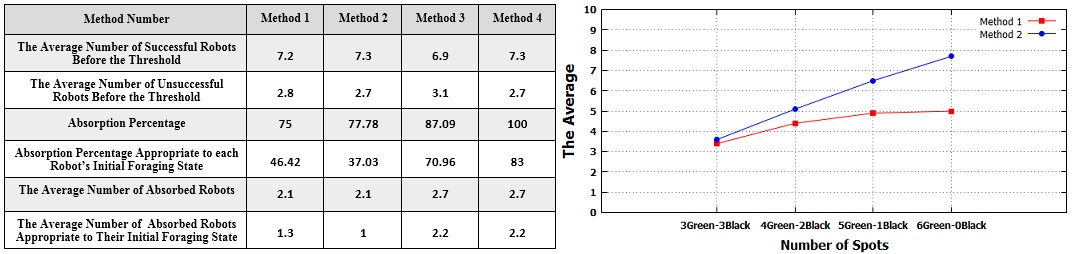}
\centering
\caption{Left:Results for the second experiment.Right:Average number of robots deployed in green spots.}
\label{fig:example}
\end{figure}

With respect to robustness, the first and second methods are completely robust against single point of failure because they behave in a distributed and autonomous manner. If a robot fails outside a spot, others will consider it as an obstacle. If it fails inside a spot and before announcing the center of it, there is no problem because it looks like the situation in which the spot is not discovered yet. But if it fails after the center announcement, it causes only decreasing in speed of the operations in the spot, because they can be carried out by other active robots on the spot yet.

The third method is less robust. The strength of the method for worker robots is such as the first and second methods. In the case of Chapar robots failures, there will be not any problem because in the worst case the speed of communication will be reduced and other robots largely compensate this loss. But about Chapar station as we have mentioned previously in describing the method to benefit the high-speed informing relative to the cost, it is necessary that their communication radius cover each other sequentially. Thus, if one of Chapar stations fails, the connection between two parts of the environment will be lost. Of course this problem can be solved by putting more than two Chapar stations in their each other communication ranges, but in this way costs will increase. It can be said that this method is robust against single point of failure. But this robustness is less than the first two methods (with respect to maintaining the efficiency of the method). Centralized Chapar method has the lowest robustness than the previous three methods because this method is only have a wide range Chapar station. So if this Chapar station fails for any reason, the efficiency of the method will be reduced so much.
\section{Conclusions}
In this paper a new practical scenario in swarm robotics is presented which is about task allocation in unknown environments. Here we consider that there is limited source of energy for each Robot. It is pointed out that energy management in the form of a 3 level structure  is essential and four self-organized threshold-based methods are proposed for solving the scenario. 

Moreover, in two general experiments, the performance of them is analyzed. From the results of simulations we can conclude that the second method is more efficient in comparison to the first one before the threshold and so the performance of the third and the fourth methods will increase by applying it as their worker robot's algorithm. However, since the methods are using similar communication mechanisms, they have almost the same performance after the threshold. 

As a result, from performance point of view, the fourth method has the highest performance, followed by the third method which in turn, has higher rank than the two others. The second method also is better than the first one in this criteria. On the other hand, from scalability and robustness perspective, both first and second methods have higher rank than the two others and further more the third method has higher rank than the fourth.

\end{document}